\documentclass[aps,prl,reprint,amsmath,showpacs,superscriptaddress, longbibliography]{revtex4-1}
\usepackage{graphicx}
\usepackage{textcomp}
\usepackage{float}
\usepackage{xcolor}
\usepackage[normalem]{ulem}
\usepackage{upgreek}
\usepackage{siunitx}
\usepackage{changes}

\newcommand{\wTA}{\omega_\mathrm{TA}}
\newcommand{\wG}{\omega_\mathrm{G}}

\newcommand{\WTA}{\Gamma_\mathrm{TA}}

\newcommand{\WD}{\Gamma_\mathrm{2D}}

\newcommand{\mum}{\mathrm{\upmu m}}

\newcommand{\cm}{\mathrm{cm^{-1}}}

\begin{document}

\title{
Raman imaging of twist angle variations in twisted bilayer graphene at intermediate angles
}

\author{A.~Sch\"apers}
    \thanks{These authors contributed equally.}
    \affiliation{JARA-FIT and 2nd Institute of Physics, RWTH Aachen University, 52074 Aachen, Germany}

\author{J.~Sonntag}
    \thanks{These authors contributed equally.}
    \affiliation{JARA-FIT and 2nd Institute of Physics, RWTH Aachen University, 52074 Aachen, Germany}
    \affiliation{Peter Gr{\"u}nberg Institute (PGI-9), Forschungszentrum J{\"u}lich, 52425 J{\"u}lich, Germany}

\author{L.~Valerius}
    \affiliation{JARA-FIT and 2nd Institute of Physics, RWTH Aachen University, 52074 Aachen, Germany}

\author{B.~Pestka}
    \affiliation{2nd Institute of Physics B and JARA-FIT, RWTH Aachen University, 52074 Aachen, Germany}

\author{J.~Strasdas}
    \affiliation{2nd Institute of Physics B and JARA-FIT, RWTH Aachen University, 52074 Aachen, Germany}

\author{K.~Watanabe}
    \affiliation{Research Center for Functional Materials,
National Institute for Materials Science, 1-1 Namiki, Tsukuba 305-0044, Japan}

\author{T.~Taniguchi}
    \affiliation{International Center for Materials Nanoarchitectonics,
National Institute for Materials Science,  1-1 Namiki, Tsukuba 305-0044, Japan}%

 \author{L.~Wirtz}
     \affiliation{Department of Physics and Materials Science, University of Luxembourg, 162a avenue de la Fa\"iencerie, L-1511 Luxembourg, Luxembourg}

 \author{M.~Morgenstern}
     \affiliation{2nd Institute of Physics B and JARA-FIT, RWTH Aachen University, 52074 Aachen, Germany}

 \author{B.~Beschoten}
     \email{Corresponding author: bernd.beschoten@physik.rwth-aachen.de}
     \affiliation{JARA-FIT and 2nd Institute of Physics, RWTH Aachen University, 52074 Aachen, Germany}

\author{R.~J.~Dolleman}
     \affiliation{JARA-FIT and 2nd Institute of Physics, RWTH Aachen University, 52074 Aachen, Germany}

\author{C.~Stampfer}
    \affiliation{JARA-FIT and 2nd Institute of Physics, RWTH Aachen University, 52074 Aachen, Germany}
    \affiliation{Peter Gr{\"u}nberg Institute (PGI-9), Forschungszentrum J{\"u}lich, 52425 J{\"u}lich, Germany}

\date{\today}

\begin{abstract}
	Van der Waals layered materials with well-defined twist angles between the crystal lattices of individual layers have attracted increasing attention due to the emergence of unexpected material properties.
	As many properties critically depend on the exact twist angle and its spatial homogeneity, there is a need for a fast and non-invasive characterization technique of the local twist angle, to be applied preferably right after stacking.
	We demonstrate that confocal Raman spectroscopy can be utilized to spatially map the twist angle in stacked bilayer graphene for angles between \SI{6.5}{\degree} and \SI{8}{\degree} when using a green excitation laser.
	The twist angles can directly be extracted from the
	moiré superlattice-activated Raman scattering process of the transverse acoustic (TA) phonon mode.
	Furthermore, we show that the width of the TA Raman peak contains valuable information on spatial twist angle variations on length scales below the laser spot size of $\sim \SI{500}{\nano\meter}$.
\end{abstract}

\maketitle

A unique degree of freedom in stacked two-dimensional materials is the twist angle, $\theta$, between the crystal lattice of adjacent layers.
The interlayer twist angle leads to the formation of superstructures with large periodicity, so-called moiré patterns, which have been shown to strongly influence the electronic, optical and phononic properties of twisted homo- and heterostructures~\cite{Wang2020Aug,Yin2015Nov,Luican2011Mar,Lin2018Aug,Huang2021Feb}.
In particular, structures of twisted graphene layers show a wide range of interesting
phenomena including superconductivity, ferromagnetism, Fermi velocity reduction or the formation of topological channels~\cite{Yin2015Nov,Luican2011Mar,Cao2018,Cao2018Apr,Liu2020Jul,Rickhaus2018Nov,Rickhaus2019Dec,Rickhaus2020Mar,deVries2020Oct,Lu2019Oct}.
Many of these phenomena can be explored in twisted bilayer graphene near the so-called magic angle, around $\theta \sim \SI{1.1}{\degree}$.
Larger twist angles enable a fine control over the Fermi velocity~\cite{Nishi2017Feb} and the tailoring of optoelectronic properties~\cite{Yin2016Mar,Pogna2021Jan}.
For example, tunable van Hove singularities allow for twist angle controlled visible and near-infrared absorption enhancement, making twisted bilayer graphene interesting for sensitive wavelength selective photo-detectors~\cite{Yin2016Mar}.
All these effects not only strongly depend on $\theta$, but are further affected by spatial variations and local gradients in $\theta$, which may occur during the sample fabrication and have been shown to yield strong unscreened in-plane electric fields or change the local band structure~\cite{Uri2020May}.
It is therefore important to establish non-invasive characterization techniques for the spatial probing of twist angles, which can be used during device fabrication and help to link physical phenomena to twist angles and spatial twist angle variations.

Different experimental techniques have been employed to spatially resolve variations in $\theta$, such as (1) transmission electron microscopy~\cite{Alden2013,Yoo2019May}, (2) atomic force microscopy (AFM)~\cite{Gallagher2016Feb,McGilly2020Jul,Sunku2018Dec}, (3) scanning tunneling spectroscopy~\cite{Li2010Feb,Yankowitz2012,Kerelsky2019Aug,Choi2019Nov,Xie2019Aug,Jiang2019Sep,Hattendorf2013},
(4) tip-enhanced Raman spectroscopy~\cite{Gadelha2021Feb},
(5) azimuthal scanning electron microscopy~\cite{Sushko2019Dec}, (6) scanning microwave impedance microscopy~\cite{Lee2020Dec}, and (7) scanning SQUID microscopy~\cite{Uri2020May}.
However, these techniques are either not applicable for tBLG fully encapsulated in hexagonal boron nitride (hBN) (1-4), lack $\theta$ sensitivity (5,6), have low throughput (1,3,4,7), require low temperatures (7) or specialized sample preparation (1,3,4,7).
In contrast, confocal Raman spectroscopy, as a promising optical characterization technique to determine $\theta$, is fast, spatially-resolved, non-invasive and allows to probe buried layers.
This method has been recently employed to image $\theta$ in twisted transition-metal dichalcogenides~\cite{Lin2020Dec}.
Its capabilities to resolve $\theta$ in graphene are however limited, when considering {the energy, i.e. frequency of} the major G{, D} and 2D Raman modes of graphene, due to their weak dependence on $\theta$~{\cite{Kim2012Jun,Gadelha2021Feb,gupta2010nondispersive,Havener2012Jun}}.

\begin{figure*}
	\includegraphics[width=\linewidth]{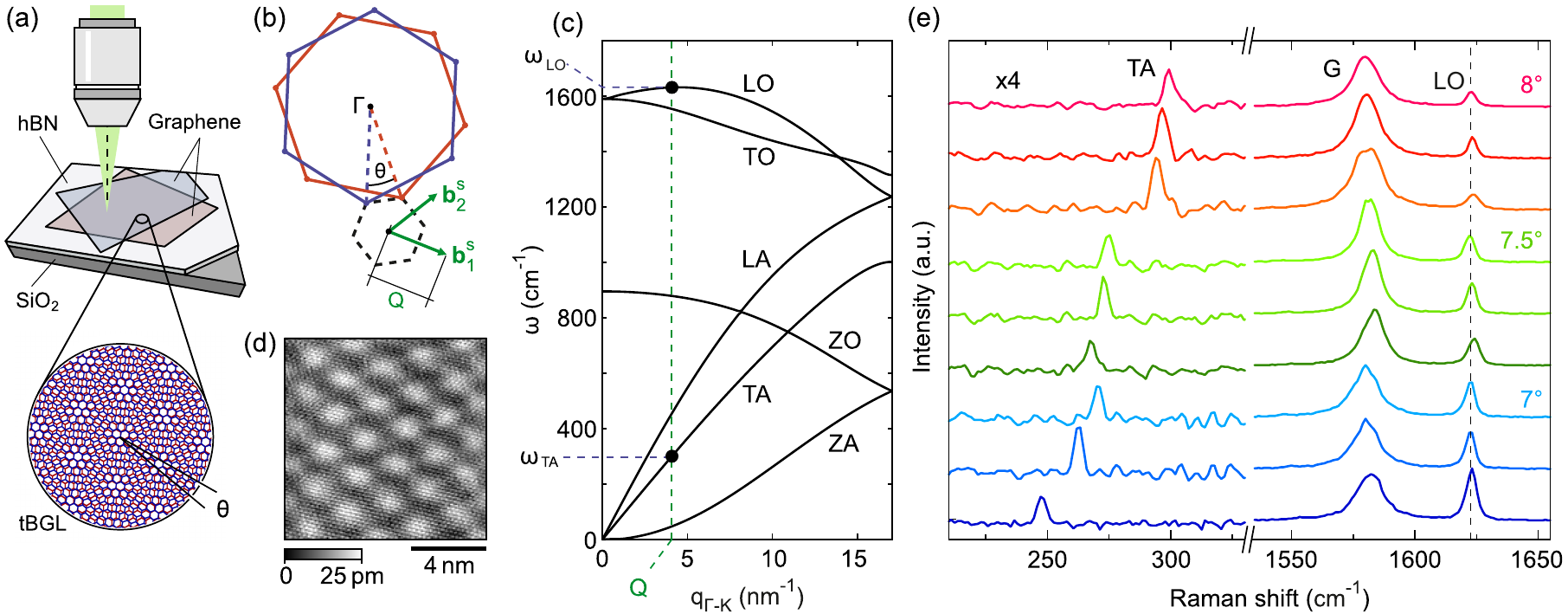}
	\caption{(a) Schematics of the experimental setup for spatial Raman mapping of tBLG on hBN. The lower inset highlights the formed moiré pattern for a given twist angle $\theta$. (b) A mini Brillouin zone with reciprocal lattice vectors $\mathbf{b}_{1,2}^\mathrm{s}$ emerges due to the additional periodicity of the superlattice.  (c) Phonon dispersion along the $\Gamma$-K direction highlighting the three acoustic (A) and three optical (O) phonon branches.
	The vertical green dotted line indicates the momentum $Q$ provided by the superlattice, which allows additional Raman scattering.
	(d) Scanning tunneling microscopy image with atomic resolution of fabricated tBLG structure with a twist angle of \SI{7.5}{\degree} as deduced from the moir\'{e} periodicity.
	(e) Raman spectra taken on three different tBLG samples with different nominal twist angels: \SI{8}{\degree} (red/orange traces), \SI{7.5}{\degree} (green) and \SI{7}{\degree} (blue). For each sample 3 spectra on different positions are shown highlighting the different twist angle sensitivity of the superlattice-induced Raman peaks (TA and LO). Here, the Raman signal of the silicon background is subtracted for clarity. }
	\label{fig01}
\end{figure*}

Here we utilize superlattice-activated Raman scattering processes leading to additional Raman peaks that have been shown to exhibit a large energy dependence on the twist angle {\cite{Campos-Delgado2013Apr,Jorio2013Dec,Carozo2011Nov,Righi2011Umklapp,righi2013resonance,wang2013resonance}} and thus promise a {high precision in determining $\theta$}.
By performing spatially-resolved confocal Raman spectroscopy on twisted bilayer graphene we show that the superlattice-induced transverse acoustic (TA) Raman peak at $\wTA\approx \SI{275}{\per\centi\meter}$ can be used to spatially map the twist angles with a {precision to resolve changes in $\theta$ better than $0.01^{\circ}$ } for twist angles ranging from $\SI{6.5}{\degree}$ to $\SI{8}{\degree}$.
This high precision enables us to spatially resolve minute twist angle variations within the device, making confocal Raman mapping a very valuable method for assessing the twist angle homogenity.
Furthermore, we show that the observed TA peak width, $\WTA$, can be used as a measure of nanometer-scale twist angle variations {smaller than the size of the laser spot}, {which allows} for the identification of regions {with a homogeneous twist angle}.

We start by recalling the physical processes behind the twist angle-related Raman peaks in stacked bilayer graphene~\cite{He2013Aug,Jorio2013Dec,Carozo2011Nov,Campos-Delgado2013Apr}.
When two graphene layers are stacked with a twist angle $\theta$ between their lattices,
a superlattice is formed in real space (see lattice structure in Figure~\ref{fig01}a).
In reciprocal space, this additional periodicity results in a new mini-Brillouin zone (dotted lines in Figure~\ref{fig01}b) with lattice vectors {$\mathbf{b}_{1,2}^\mathrm{s}(\theta)$ } of length
\begin{equation}
   Q(\theta)=
   {|\mathbf{b}_{1,2}^\mathrm{s}(\theta)|}
   =\frac{8\pi}{\sqrt{3}a}\mathrm{sin}\left(\frac{\theta}{2}\right),
     \label{eq:qtheta}
\end{equation}
where $a=\SI{0.247}{\nano\meter}$ is the lattice parameter of graphene.
This reduction in Brillouin zone size can be viewed as the ability of the superlattice to provide momenta of multiples of its reciprocal lattice vectors {$\mathbf{b}_{1,2}^\mathrm{s}(\theta)$}.
As a result, the momentum selection rules for Raman scattering allow probing of additional phonon modes that are located at {$\Gamma\pm\mathbf{b}_{1,2}^\mathrm{s}(\theta)$}, i.e. these modes become Raman active~\cite{He2013Aug,Jorio2013Dec,Carozo2011Nov,Campos-Delgado2013Apr}.
The respective frequencies of these superlattice-induced Raman peaks depend on $\theta$ due to the dispersion of the phonon modes (see Figure~\ref{fig01}c) and the relation between $Q$ and $\theta$ (eq 1).
{Although the path in reciprocal space defined by $\mathbf{b}_{1,2}^\mathrm{s}(\theta)$ slightly deviates from the path along $\Gamma$-K, we use the phonon dispersion along this high-symmetry axis in our 
analysis.}
{The accuracy of this approximation is discussed below and in the Supplementary Information.}
It is important to emphasize that these Raman peaks result from resonant Raman scattering processes, i.e. their respective intensities depend on the combination of the twist angle and laser energy used for the Raman experiments

The investigated van der Waals heterostructures were built using the tear and stack method~\cite{Kim2016Mar,Cao2016Sep} and consist of an hBN crystal (\SIrange{20}{30}{\nano\meter} thick) with tBLG of the respective nominal twist angle on top (see Figure~\ref{fig01}a).
To achieve this stacking orientation, the heterostructures were {fabricated using a polydimethylsiloxane/polyvinyl alcohol (PDMS/PVA) stamp to first pick up a hBN flake followed by two graphene sheets torn from the same flake. In the next step, the hBN/tBLG heterostructure is flipped by depositing it on a PDMS/poly(bisphenol A carbonate) (PDMS/PC) stamp with the tBLG lying on the polymer. After this step, the heterostructure is transferred to a Si$^{++}$/SiO$_2$ substrate and the tBLG is lying on top of the hBN crystal.} {The fabrication procedure is described in detail in the Supplemental Information.}
Prior to the Raman measurements, we employed an AFM cleaning step~\cite{Goossens2012,Kim2019Dec} to remove residual polymers from the surface of the tBLG.
In Figure~\ref{fig01}d we show a scanning tunneling microscopy image of such a tBLG sample{, demonstrating that we indeed obtain clean samples. Furthermore, a periodicity is observed that can be attributed to the formation of a moiré pattern between the graphene layers with a twist angle of $\SI{7.5}{\degree}$ \cite{Shallcross2010Turbostratic}. }

\begin{figure*}
	\includegraphics[width=\linewidth]{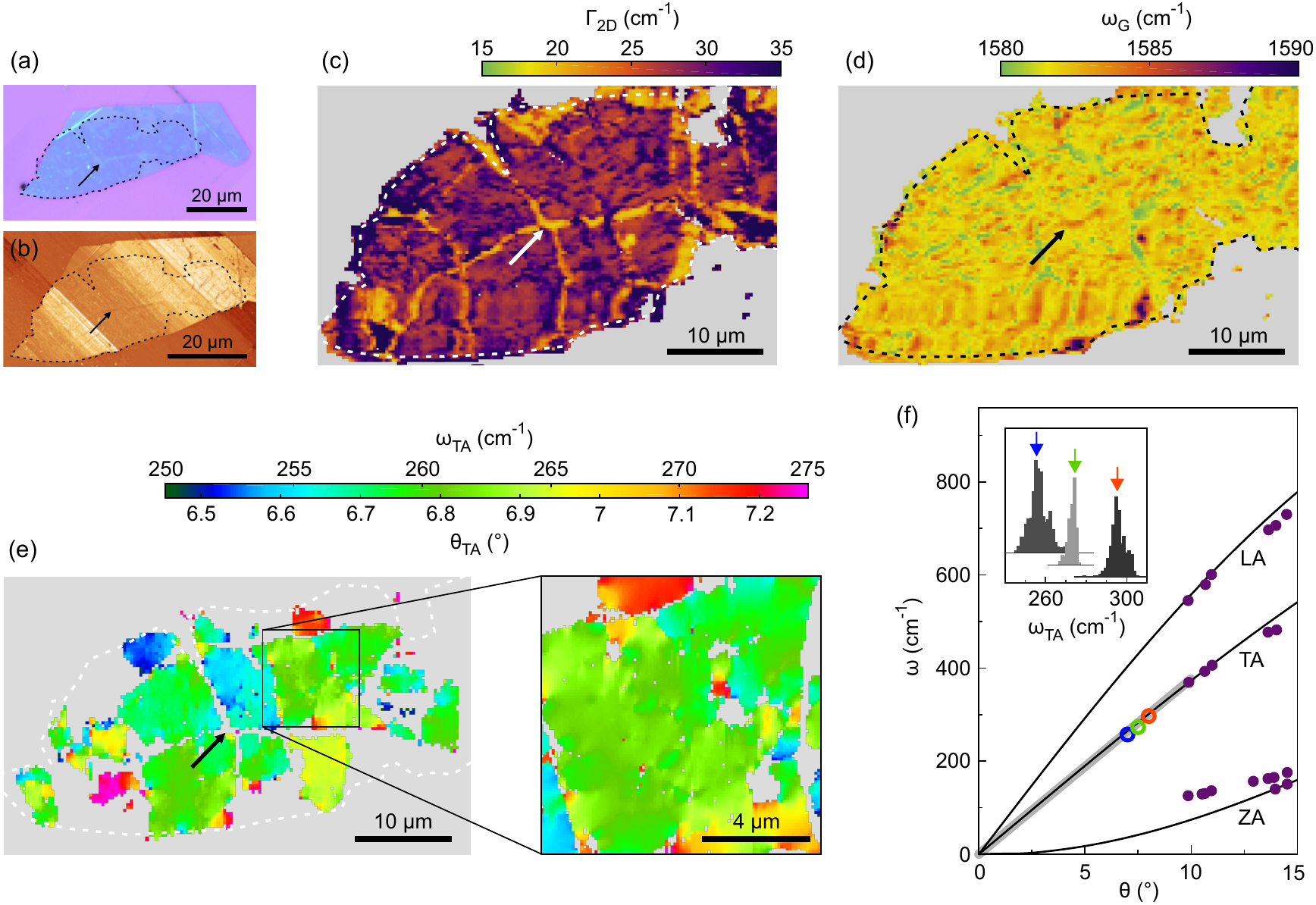}
	\caption{(a) Optical image of tBLG ($\theta_\mathrm{n} = \SI{7}{\degree}$) on hBN (large {blue} crystal). The dashed lines indicate the extend of the tBLG region. (b) Phase channel of an AFM image of the same heterostructure. {The arrow marks a ruptured area of the tBLG.} (c) The width of the Raman 2D peak $\WD$ highlights folds and cracks within one graphene layer, as indicated e.g. by the arrow. (d) The position of the Raman G peak $\wG$ shows only minor variations over the entire sample, ruling out strong doping {and} strain variations. (e) Left panel: The position of the Raman TA peak $\wTA$ shows large spatial variation and the formation of distinct domains between the cracks and folds. Right panel: Close-up of the square displayed in the left panel, showing $\wTA$ variations within one "domain".
	(f) Phonon dispersion along the $\Gamma$-K direction, where $q_{\Gamma-\mathrm{K}}$ is converted to $\theta$ via eq~\ref{eq:qtheta}. The colored circles represent the mean positions of $\wTA$ of the three samples (see inset) built with twist angles of $\theta_\mathrm{n} \approx \SI{7}{\degree},~\SI{7.5}{\degree},~\SI{8}{\degree}$. The color code corresponds to Figure~\ref{fig01}e. The purple dots are taken from Ref.~\cite{Campos-Delgado2013Apr}. The bold gray line is a linear fit to $\wTA$ for $\theta<\SI{10}{\degree}$. The inset shows histograms of  $\wTA$ extracted from the Raman maps of the three different samples (see colored arrows).}
	\label{fig02}
\end{figure*}

In our twisted bilayer graphene samples
we find that {the} superlattice-induced {TA and LO} Raman peaks appear most prominently in samples built with twist angles between $\SI{6.5}{\degree}$ to $\SI{8}{\degree}$.
{This is in line with an analysis that combines the optical resonance conditions with symmetry considerations of the electron-phonon interaction as outlined in Ref. \cite{Eliel2018intra}. For a green laser with a photon energy of \SI{2.33}{\electronvolt} and twist angles near \SI{7}{\degree}, we accordingly identify the superlattice-induced peaks as the result of an intralayer process excited near the $K$-point, where the required crystal momentum for the phonon transition matches the length of the mini-Brillouin zone $Q(\theta)$~\cite{Carozo2011Nov,Jorio2013Dec,Campos-Delgado2013Apr}. These energy and momentum considerations would in principle allow all phonon branches to become Raman-active, but due to symmetry considerations only phonons from the TA and LO branch are allowed to partake in intralayer transitions and thus become observable in our measurement \cite{moutinho2021resonance,PhysRevB.79.125426}. }
Figure~\ref{fig01}e shows several representative Raman spectra measured on samples with nominal (n) twist angles of $\theta_\mathrm{n}=\SI{7}{\degree}$,   $\SI{7.5}{\degree}$ and $\SI{8}{\degree}$.
Although both modes {can also be} observed at larger and lower twist angles with lower intensities, we restrict our study to this angle range with large peak intensities. This allows for short data acquisition times and, hence, Raman images with high spatial resolution~\cite{Carozo2011Nov}.

The most striking observation in Figure~\ref{fig01}e is the large change in the frequency of the Raman TA peak for different twist angles, compared to the Raman G and LO peak.
The strong variation of the Raman TA peak position can be traced back to the large dispersion $\partial \omega_{\rm TA}/\partial q$ of the TA branch, as shown in Figure~\ref{fig01}c.
Importantly, we also observe strong variations of $\wTA$ within each sample, e.g. up to shifts of \SI{25}{\per\centi\meter} for the sample with $\theta_\mathrm{n}=\SI{7}{\degree}$ (blue spectra in Figure~\ref{fig01}e), indicating spatial inhomogeneities in $\theta$.

\begin{figure*}
	\includegraphics[width=\linewidth]{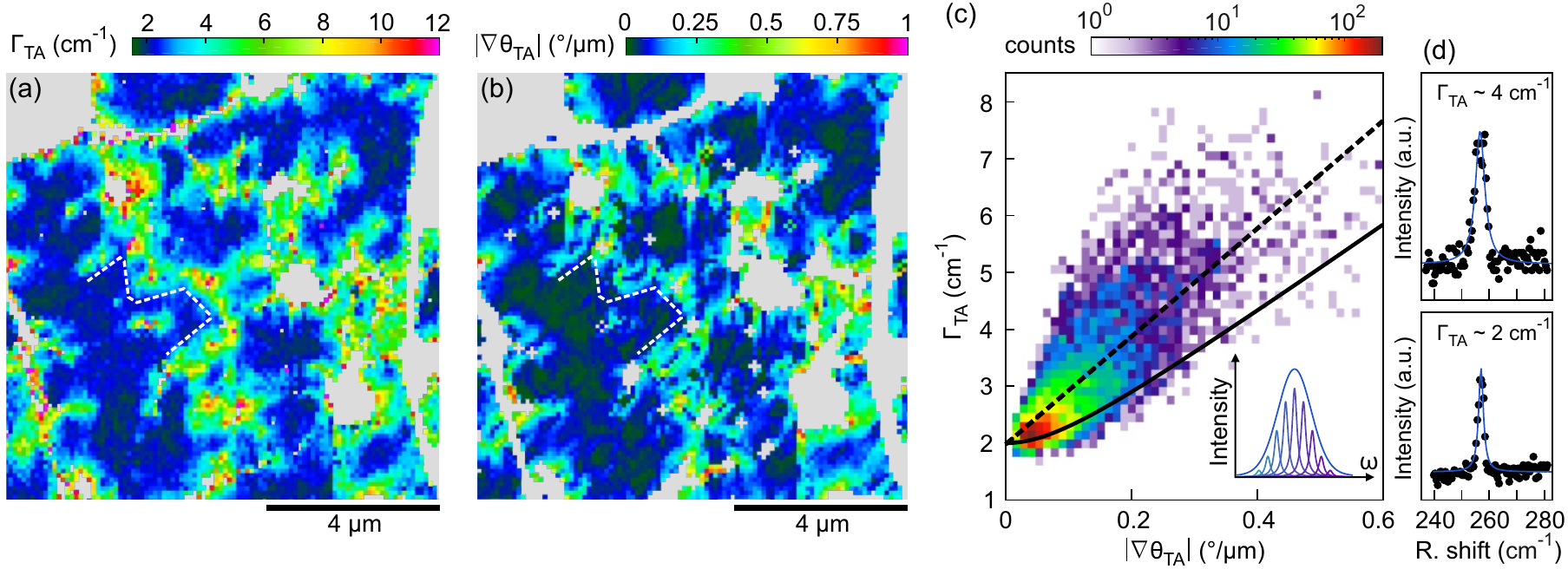}
	\caption{(a) Width of the Raman TA peak $\WTA$. (b) Numeric gradient of $\theta_\mathrm{TA}$, calculated from the close-up in Figure~\ref{fig02}e. The white dashed line in (a) and (b) is a guide to the eye to allow for easy comparison.(c) Two-dimensional histogram of the correlation between $\WTA$ and $|\nabla\theta_\mathrm{TA}|$. For clarity both values are spatially averaged over their next-nearest neighbors. The black dashed line shows a linear fit to the experimental data. The black line shows the expected broadening due to a homogeneous gradient in $\theta_\mathrm{TA}$, i.e. in $\wTA$. The inset illustrates the peak broadening effect due to $|\nabla\theta_\mathrm{TA}|$. The colored lines represent the spectral contribution of different positions within the laser spot. The pink line is the averaged Raman intensity detected at the spectrometer. (d) Two representative Raman spectra of the TA mode with respective Lorentzian fits.
	}
	\label{fig03}
\end{figure*}

For mapping the twist angle and studying its spatial homogeneity we focus on the spatial dependence of $\wTA$ and the strong dependence of $\wTA$ on $\theta$.
We present spatial Raman maps of the sample built with a nominal twist angle of $\theta_\mathrm{n} = \SI{7}{\degree}$, which shows the strongest $\theta$ variations in Figure~\ref{fig01}e.
In Figures~\ref{fig02}a and \ref{fig02}b we show an optical image and a phase image taken by atomic force microscopy, respectively, showing the finished tBLG sample with $\theta_\mathrm{n} = \SI{7}{\degree}$.
The dotted line outlines the region of tBLG.
As indicated by the arrow, the stack got ruptured along several directions during fabrication.
The ruptures are well visible in the Raman map of the 2D peak width $\WD$ in Figure~\ref{fig02}c.
The 2D peak width is known to be strongly influenced by the electronic band-structure-changes between graphene and tBLG~\cite{Kim2012Jun,Graf2007Feb}, which results in an overall narrow line width for (effective) single-layer graphene and a broader line width for bilayer graphene.
The small line width and a decrease in G peak intensity (not displayed) along the ruptures (yellowish color) shows the presence of single-layer graphene, indicating that the ruptures and cracks are predominantly present only in one of the graphene sheets in the tBLG sample.
However, the Raman G peak position $\wG$ in Figure~\ref{fig02}d is spatially quite homogeneous, indicating that there are no large amounts of residual strain or doping within the heterostructure~\cite{Mohiuddin2009May,Neumann2015b,Stampfer2007}.

In contrast, $\wTA$ (Figure~\ref{fig02}e) varies significantly and forms distinct domains mostly delimited by ruptures and cracks.
Thanks to the homogeneity in $\wG$, we exclude strain and doping as the origin of these variations in $\wTA$ and therefore conclude that variations in $\wTA$ result from changes in the twist angle $\theta$.
We note that the $\wTA$ mode cannot be extracted for all regions of the tBLG.
This might either be due to disordered folding of the graphene sheets near the edges or due to local twist angle values beyond the accessible range.
We next convert the $\wTA$ values to twist angles $\theta$ by inverting eq.~\ref{eq:qtheta} and using the phonon band structure of graphene.
To verify this relation, we first compare the mean values of $\wTA$ obtained from the three samples to the phonon dispersion along $\Gamma$-K, where $q_{\Gamma-\mathrm{K}}$ is converted to a twist angle  $\theta$.
As shown in Figure~\ref{fig02}f, we find a good agreement between the mean $\wTA$ of our three samples and the phonon dispersion, when using $\theta_\mathrm{n} = \SI{7}{\degree}, \SI{7.5}{\degree}$ and $\SI{8}{\degree}$ as targeted during fabrication.
This finding is consistent with previous studies~\cite{Campos-Delgado2013Apr} (see purple dots in Figure~\ref{fig02}f) and verifies the $\theta$-dependence of $\wTA$, which allows to invert $\wTA(\theta)$ and thus directly translate the $\wTA$ into $\theta_\mathrm{TA}$.
The index of $\theta_\mathrm{TA}$ indicates that the twist angle is determined via the $\wTA$ to $\theta$ relation in Figure~\ref{fig02}f, in contrast to the nominal twist angle $\theta_\mathrm{n}$ defined during fabrication.
In Figure~\ref{fig02}e we directly converted the $\wTA$ scale bar into a $\theta_\mathrm{TA}$ scale bar {using the theoretical phonon dispersion}.
It shows that even within the same tBLG stack aimed at $\theta_\mathrm{n}=7^\circ$, there are large spatial variations in $\theta_\mathrm{TA}$ ranging from $\SI{6.5}{\degree}$ to $\SI{7.3}{\degree}$.

To estimate the {accuracy and precision to determine the} twist angle by this technique, we linearize $\wTA(\theta)$ in this angle regime (see bold gray line in Figure~\ref{fig02}f) to obtain a simple expression of the twist angle given by
$\theta_{_\mathrm{TA}} \approx 0.0265\mathrm{^\circ}\cdot\wTA /\mathrm{cm}^{-1}$. 
{By comparing our theoretical dispersion with earlier experiments~\cite{cong2019probing,Oshima81,Siebentritt97,Yanagisawa05} (see Supplementary Information), we estimate that there is an uncertainty of $\approx 9$\% on the slope of $0.0265\mathrm{^\circ}/\mathrm{cm}^{-1}$. This translates into a $\approx 9$\% uncertainty on the \emph{absolute} value of the twist angle in Figure~\ref{fig02}e.}
{To determine the precision of our method, which defines our ability to resolve \emph{changes} in the twist angle as a function of position, we analyze all the fitting results for the map in Figure~\ref{fig02}e and obtain an average 95\% confidence interval for the peak position of $\approx$\SI{0.25}{\per\centi\meter}. Using the expression above and taking into account the uncertainty in the slope, we expect that our precision is better than \SI{0.01}{\degree}.
This high precision allows us to map minute twist angle variations within the sample as shown in Figure~\ref{fig02}e, which is a key result of this work.}

Twist angle variations not only occur near ruptures, cracks or folds in the heterostructure, but also within otherwise continuous regions of tBLG, as illustrated by the close-up shown in Figure~\ref{fig02}e.
This is in agreement with previous findings using different characterization techniques~\cite{Uri2020May,Sushko2019Dec,Lu2019Oct,Lin2020Dec,Lee2020Dec},
but makes it difficult to directly compare twist angle values extracted from Raman spectroscopy data and from STM images. This is mainly because of difficulties in identifying the exact same locations with both methods.
It is worth noting that in one case investigated, good agreement was found within the margins of error (for more details, see Supplementary Information).
The information on $\theta_\mathrm{TA}$ obtained via confocal Raman spectroscopy is suitable for a fast characterization of tBLG stacks, or for the preselection of high-quality areas with high $\theta$ homogeneity for subsequent device fabrication.

{The spatial resolution of this method is, however, limited by the laser spot size, making it difficult to resolve twist angle variations smaller than the spot size. }
{To obtain further insight into twist angle variations on length scales smaller than the laser spot size, we can analyse the width of the Raman TA peak, $\WTA$.}
Figure~\ref{fig03}a shows a map of $\WTA$ corresponding to the magnified area shown in Figure~\ref{fig02}e.
We note that these measurements were taken with a grating of 2400 lines/mm to allow for the high energy resolution needed for the line width analysis.
As can be seen in Figure~\ref{fig03}a, there are substantial variations in $\WTA$ extending over a range of \SIrange{2}{10}{\per\centi\meter} (see also two representative Raman spectra with respective Lorentzian fits in Figure~\ref{fig03}d).
The largest line widths are observed in areas of large variations in $\wTA$, i.e. in $\theta_\mathrm{TA}$ (compare to Figure~\ref{fig02}e).
We highlight this by numerically calculating the magnitude of the gradient $|\nabla\theta_\mathrm{TA}|$ as shown in Figure~\ref{fig03}b (for an easy comparison see white dashed lines in Figures~\ref{fig03}a and \ref{fig03}b).
To evaluate the correlation in more detail, we plot $\WTA$ versus $|\nabla\theta_\mathrm{TA}|$ in Figure~\ref{fig03}c, where we employed next-nearest neighbor averaging.
The correlation map suggests that the increase in $\WTA$ in areas of large $|\nabla\theta_\mathrm{TA}|$ is caused by locally averaging various spectra of different $\theta_\mathrm{TA}$, i.e. $\wTA$, within the laser spot.
This effect is similar to the known strain variation-induced broadening of the 2D Raman peak in graphene~\cite{Neumann2015b,graphenestandard}.
The inset of Figure~\ref{fig03}c illustrates how this statistical averaging effect broadens the TA peak.

To further verify that the peak broadening can be attributed to statistical averaging, we perform a numerical estimate of the expected $\WTA$ for a given gradient $|\nabla\theta_\mathrm{TA}|$.
This is done by convolving the Gaussian shape of the laser spot with a Lorentzian peak describing the Raman TA mode with a linearly varying $\wTA$, which corresponds to a constant $|\nabla\theta_\mathrm{TA}|$~\cite{Neumann2015b}.
The Gaussian peak is assumed to have a full width at half maximum (FWHM) of \SI{520}{\nano\meter} (corresponding to the measured laser spot size, as shown in the Supplementary Information), and we use the lowest experimentally measured $\WTA$ as the intrinsic width $\WTA(|\nabla\theta| \approx 0) \approx \SI{2}{\per\centi\meter}$.
After this convolution, we use a Lorentzian fit to determine $\WTA$ of the convolved peak.
The result is shown as black line in Figure~\ref{fig03}c, representing the Raman TA peak width if the gradient $|\nabla\theta_\mathrm{TA}|$ is homogeneous within the laser spot.

Since any inhomogeneities would lead to a TA peak broadening, we expect this line to represent a lower bound which agrees well with the experimental data in Figure~\ref{fig03}c.
Data points close to the black line represent areas of the tBLG where the broadening $\WTA$ is dominated by twist angle variations extending to length scales larger than the laser spot size (\SI{520}{\nano\meter}).
Values of $\WTA$ above this line correspond to areas with large variations in $\theta$ on length scales smaller than the spot size, which cannot be spatially-resolved and thus do not contribute to $|\nabla\theta_\mathrm{TA}|$.
To quantify the empirically found correlation of $\WTA$ and $|\nabla\theta_\mathrm{TA}|$ in Figure~\ref{fig03}c, we perform a linear fit to our experimental data as shown by the black dashed line in Figure~\ref{fig03}c.
We find $\WTA(|\nabla\theta| \approx 0)=\SI{1.97 \pm 0.01}{\per\centi\meter}$ and a slope of $\beta=\SI{9.47 \pm 0.08}{\per\centi\meter}/(^\circ/\mum)$.
The average twist angle gradient within our laser spot is thus given by
$|\overline{\nabla\theta}_\mathrm{TA}|=0.106~(^\circ/\mum)\cdot(\WTA-1.97\,\cm)/(\cm)$,
where $\WTA$ is the experimentally measured TA peak width.
This analysis demonstrates that measuring the Raman TA peak position and width allows to spatially map the twist angle and the twist angle variations in tBLG.

In our study, we have focused on tBLG with $\theta \approx 6.5^\circ$--$8^\circ$ due to the resonant effects enhancing the intensity of the TA peak.
{To characterize samples with smaller or larger twist angles, the resonance conditions have to be considered \cite{Carozo2011Nov,Jorio2013Dec,Carozo2013Resonance,Eliel2018intra}.}
{For small twist angles $\theta \lesssim  \SI{4}{\degree}$, the excitation energy of the laser can be reduced into the infrared regime.}
{Close to the magic angle, however, the electronic band structure of graphene flattens \cite{Cao2018}, which may reduce the resonance energy even further than the $\sim$0.5 eV predicted by theory \cite{Carozo2011Nov, moutinho2021resonance}}.
{{To circumvent this problem,} small twist angles may also be accessed through electron-phonon processes close to the M point, which requires larger excitation energies in the UV regime {\cite{Eliel2018intra,righi2013resonance}}.}
{Lattice relaxation effects are another important consideration at small twist angles, since these can alter the phonon dispersion \cite{PhysRevB.100.155426}. While the method can still provide qualitative information on twist angle inhomogeneities, for the absolute value of $\theta$ one needs to replace the SLG phonon dispersion with an appropriate calculation. }
{Our method can also be extended to larger angles with the 2.33 eV excitation, since the TA peak can be observed for $\theta>15^\circ$ due to an interlayer electron-phonon process close to the K~point~\cite{Campos-Delgado2013Apr,Jorio2013Dec}.}

{Our method for determining the twist angle is also interesting for benchmarking different fabrication techniques that are aimed to reduce twist angle variations. Different approaches have emerged to improve the fabrication of twisted bilayer graphene, for example by using pyramid-shaped polymer stamps \cite{gadelha2021twisted}, pre-cutting the graphene by local anodic oxidation or laser cutting \cite{li2018electrode,park2021flavour}, and mechanical cleaning by atomic force microscopy \cite{Lu2019Oct}. By constructing samples
    in the range of 6.5$^{\circ}$ to 8$^{\circ}$, our method can quantify whether certain techniques improve the twist angle uniformity in twisted bilayer graphene. This also makes our work relevant for improving the fabrication techniques to make high-quality twisted bilayer graphene near and at the magic angle. }

In conclusion, we used confocal Raman spectroscopy to spatially map the twist angle in tBLG within a range of \SI{6.5}{\degree} to \SI{8}{\degree}.
The Raman TA peak, activated by the additional periodicity in the tBLG allows probing {minute changes in} the twist angle with a {precision better than \SI{0.01}{\degree}} and a spatial resolution of $\sim \SI{500}{\nano\meter}$.
Furthermore, we have shown that $\WTA$ is a measure of the twist angle variations on length scales below the laser spot size and can be used to estimate the average twist angle gradient.
We expect that the properties of twist angle homogeneity investigated here will play a critical role in the understanding and controlling of material properties of twisted van der Waals heterostructures.
In short, our work makes spatially-resolved confocal Raman spectroscopy an important tool for the characterization of tBLG and the bench-marking of different fabrication methods.

\begin{acknowledgments}
\textbf{Acknowledgments:}
This project has received funding from the European Union’s Horizon 2020 research and innovation programme under grant agreement No. 881603 (Graphene Flagship) and from the European Research Council (ERC) under grant agreement No. 820254, under FLAG-ERA grant TATTOOS by the Deutsche Forschungsgemeinschaft (DFG, German Research Foundation) - 437214324, the Deutsche Forschungsgemeinschaft (DFG, German Research Foundation) under Germany’s Excellence Strategy - Cluster of Excellence Matter and Light for Quantum Computing (ML4Q) EXC 2004/1 - 390534769, through DFG (STA 1146/12-1), and by the Helmholtz Nano Facility~\cite{Albrecht2017}. Growth of hexagonal boron nitride crystals was supported by the Elemental Strategy Initiative conducted by the MEXT, Japan, Grant Number JPMXP0112101001, JSPS KAKENHI Grant Numbers JP20H00354 and the CREST(JPMJCR15F3), JST. {L.~W. acknowledges financial support from the Fond National de Recherche, Luxembourg via project INTER/19/ANR/13376969/ACCEPT.}
\end{acknowledgments}
~\\
\textbf{Data Availability:} The data supporting the findings are available in a Zenodo repository under accession code 10.5281/zenodo.6631559.

\bibliography{literature}

\pagebreak
\onecolumngrid
\setcounter{equation}{0}
\setcounter{figure}{0}
\setcounter{table}{0}
\makeatletter
\renewcommand{\theequation}{S\arabic{equation}}
\renewcommand{\thefigure}{S\arabic{figure}}
\renewcommand{\thesection}{S\arabic{section}}
\renewcommand{\thetable}{S\arabic{table}}

\pagebreak
\begin{center}
    \textbf{\large Supplemental Material}
\end{center}

\section{I. Details on the Sample Fabrication}
\begin{figure}[h]
    \centering
    \includegraphics{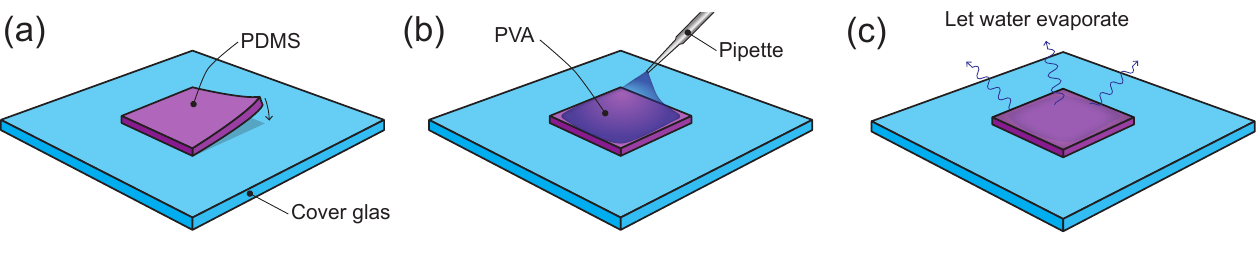}
    \caption{{Preparation of the polymer stamp. (a) A PDMS stamp is placed on the microscope cover glass. (b) A droplet of PVA solution is placed on the stamp. (c) After the water has evaporated, the stamp is ready for use.} }
    \label{fig:stamp}
\end{figure}

The fabrication technique to produce flipped half-stacks of tBLG is based on the tear and stack method, thus ensuring control over the twist angle between the graphene lattices \cite{Kim2016Mar,Cao2016Sep}.
To flip the samples, the method {is adapted} by {exploiting} the water solubility of polyvinyl alcohol (PVA) \cite{wong2020cascade}.
In the first step of fabrication, a {piece of polydimethylsiloxane (PDMS, Gelpack) approximately 5 by 5 mm is placed on a 20 by 20 mm microscope cover glass (Fig. \ref{fig:stamp}a).}
{A} small droplet of a \SI{5}{\percent} solution of PVA {(Sigma Aldrich, $M_w$ 9,000-10,000, 80\% hydrolyzed)} in water {is} dropped on the PDMS square {(Fig. \ref{fig:stamp}b)}, leaving behind a thin layer of PVA after the water {has} evaporated {(Fig. \ref{fig:stamp}c)}.
For the further process, it is important that the PVA does not reach over the edge of the PDMS.

\begin{figure}[h]
    \centering
    \includegraphics{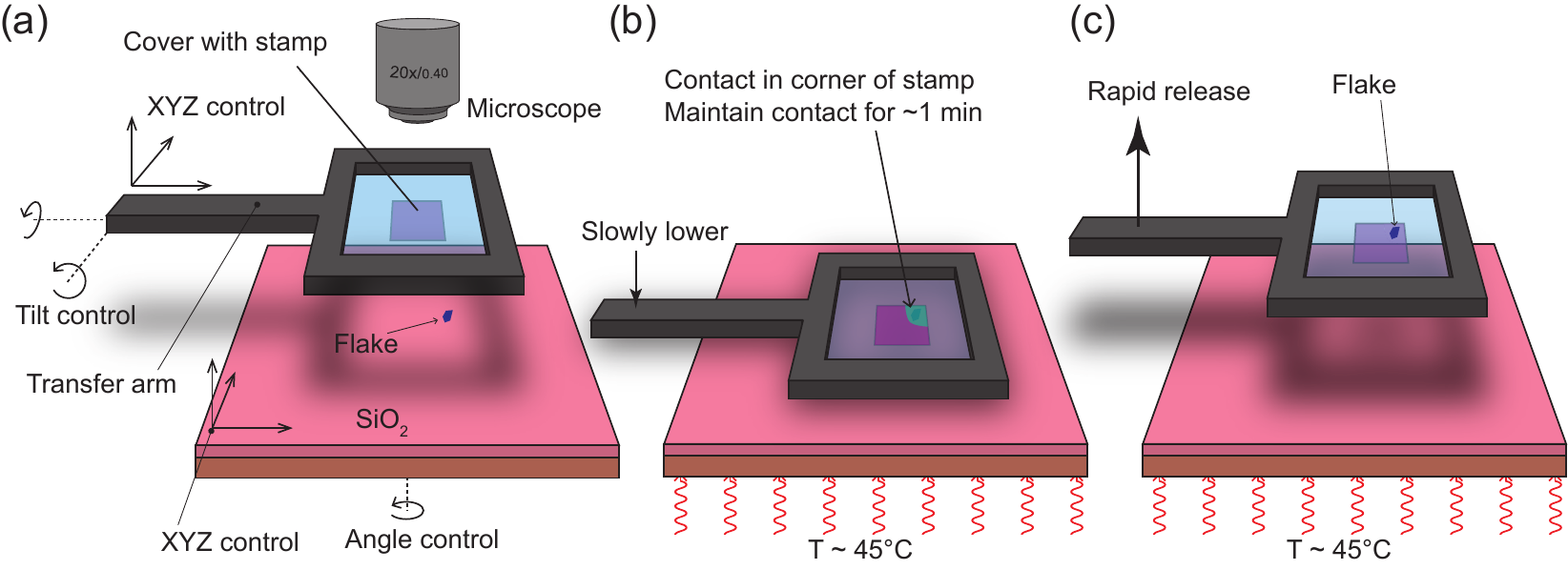}
    \caption{{Pick-up of exfoliated flakes. (a) The stamp is placed upside-down on a arm connected to a micromanipulator, and held in place by a vacuum. (b) The stamp is slowly brought into contact with the flake while the substrate is heated to 45$^{\circ}$. (c) After one minute, the flake is picked up by rapidly moving the arm upward. }}
    \label{fig:pickup}
\end{figure}

We use mechanically exfoliated hBN and graphene flakes on \SI{90}{\nano\meter} of SiO\textsubscript{2}/Si{$^{++}$} dies{, which are placed on a heated stage that is able to rotate (Fig. \ref{fig:pickup}a).}

\begin{figure}[h]
    \centering
    \includegraphics{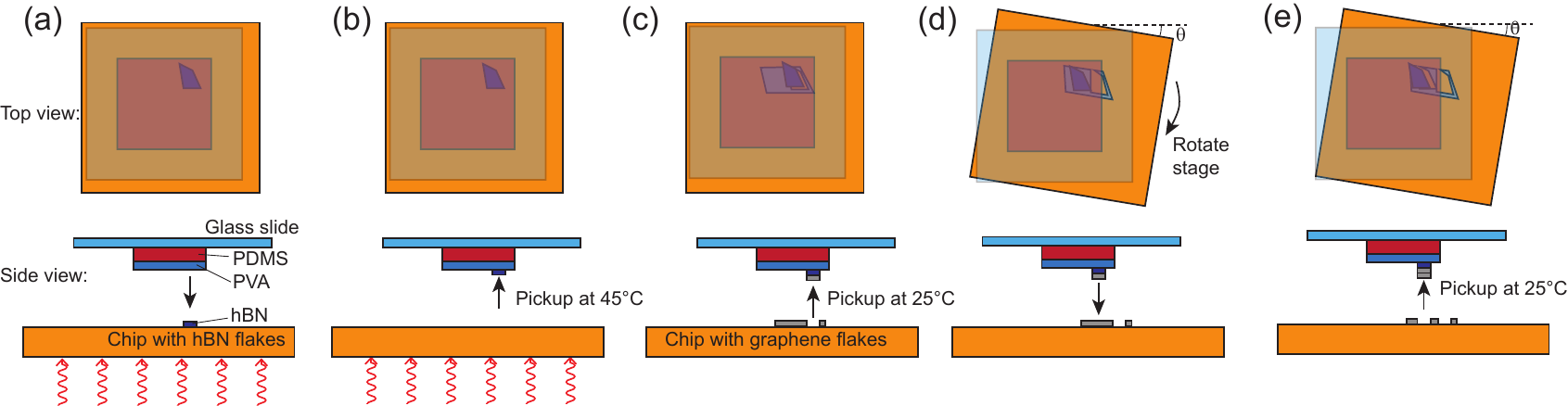}
    \caption{{Tear and stack method to assemble the twisted bilayer graphene heterostructure on the stamp.} }
    \label{fig:tearstack}
\end{figure}

{The PMDS/PVA stamp is placed upside-down in a setup with a micromanipulator, to pick up the flakes from substrate as shown in Fig. \ref{fig:pickup}. The hBN flake is picked up first, followed by the tear-and-stack procedure to pick up the graphene flakes with the desired twist angle, as illustrated in Fig. \ref{fig:tearstack}. }

\begin{figure}[h]
    \centering
    \includegraphics{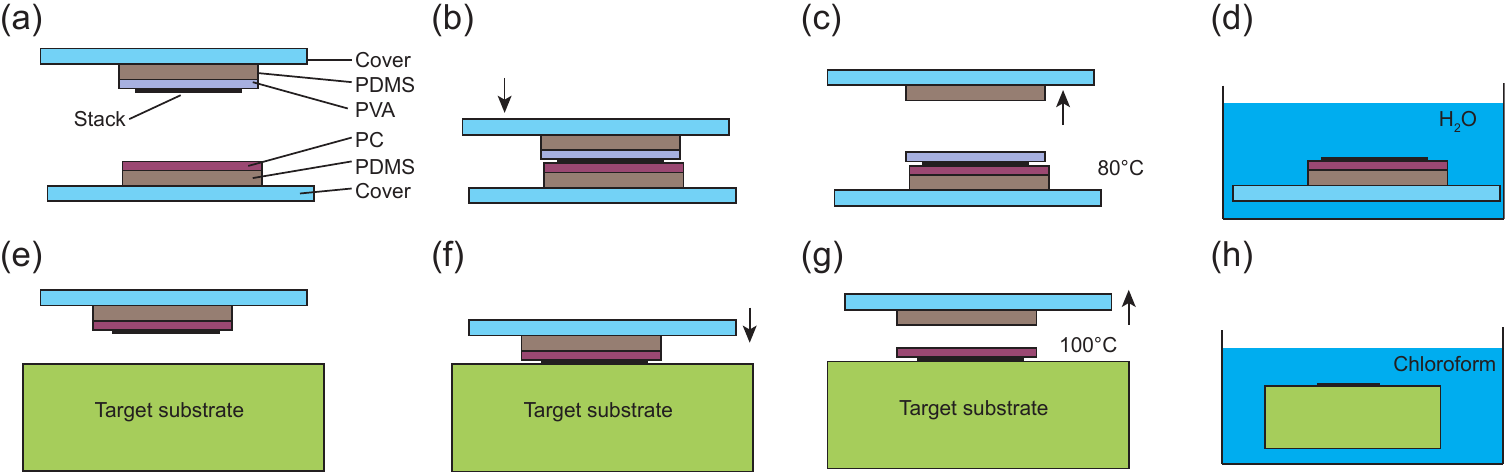}
    \caption{{Flipping the assembled stack and transferring to the final target substrate.}}
    \label{fig:flipping}
\end{figure}

When the stack {is} assembled, the order of materials is as follows: tBLG - hBN - PVA- PDMS - glass slide.
{Now, a second polymer stamp is prepared on a glass slide, consisting of a square of PDMS with a long and rectangular strip of thin poly(bisphenol A carbonate) (PC) laying across it.
The two outer ends of the PC strip are hanging over the edge of the PDMS square and they are taped to the glass slide.
Also, the area of PDMS covered by the PC strip has to be larger than the patch of PVA. }
In the next step, the layer of PVA with the sample on it is transferred from the first polymer stamp to {the second stamp with PC}.
Using the transfer system, the PVA stamp is placed on top of the PC stamp at \SI{80}{\celsius} {(Figs. \ref{fig:flipping}a-b)}. It is important to place the patch of PVA exactly on that area of PC which covers the PDMS.
The high temperature ensures that when removing the upper stamp, the PVA will detach from the PDMS {(Fig. \ref{fig:flipping}c)}, leaving the following order of materials from up to down on the PC stamp: PVA - hBN - tBLG - PC - PDMS - glass slide.
To remove the PVA from the top of the stamp, the entire glass slide is put into water for at least ten minutes {(Fig. \ref{fig:flipping}d)}. After this, the hBN - tBLG sample is again located at the surface of the stamp.
Finally, the tapes fixing the outer edges of the PC strip to the glass are removed. Alternatively, the PC strip can be cut in a way that leaves the central piece of PC covering the PDMS free to be transferred, without damaging the sample.
At a temperature of roughly \SI{100}{\celsius}, the hBN - tBLG - PC part of the stamp is transferred onto a SiO\textsubscript{2}/Si{$^{++}$} chip {(Figs. \ref{fig:flipping}e-g)}. After removing the PC with chloroform {(Fig. \ref{fig:flipping}h)}, the flipped half-sandwich of tBLG on hBN is ready.

\section{II. Details on Raman spectroscopy measurements}

\begin{figure*}[h!]
	\includegraphics{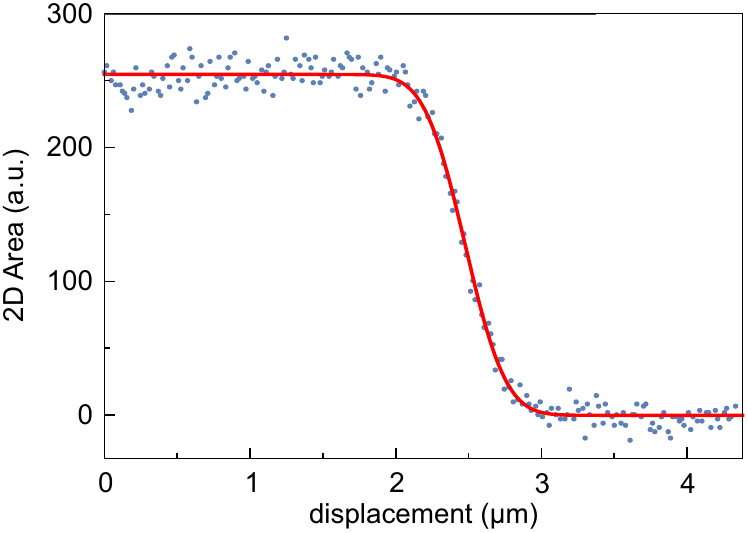}
	\caption{Intensity of the 2D Raman peak along along a path perpendicular to the edge of a graphene flake. As the edge of the flake is atomically sharp, the intensity profile is a convolution of a step function with a Gaussian function. We thus determine the laser spot size by fitting the intensity with an error function $I(x) = I_0 \textrm{erf}(\tfrac{(x-x_0)}{r})$, in which $r$ is the radius of the Gaussian profile and $x_0$ is the position of the edge. The result of the fit results in a laser spot FWHM of \SI{518 \pm 14}{\nano\meter}.}
	\label{figS01}
\end{figure*}

The Raman measurements are performed in a commercial micro-Raman setup manufactured by WITec GmbH with an excitation wavelength of $\lambda=\SI{532}{\nano\meter}$.
The laser power typically used is \SI{5}{\milli\watt}.
The laser is focused onto a sample via a \SI{100}{\times} ($\mathrm{NA}=0.9$) objective to a spot with a FWHM of $\sim \SI{520}{\nano\meter}$, see Figure~\ref{figS01}.
For the detection of the scattered light, we employ a CCD spectrometer with a grating of 1200~lines/mm when investigating  $\wTA$ or the G and 2D peak.
To extract a meaningful value of the width of the TA peak $\WTA$ we employ a higher resolution grating of 2400~lines/mm.
The higher resolution grating is used for the analysis of the width of the TA peak $\WTA$, i.e., in the measurements shown in the zoom-in of Figure~2e and Figure~3 in the main text.
The typical integration time for mapping is \SI{5}{\second}.

\section{3. Scanning tunneling Microscopy}
The flipped stacks of Si/SiO$_2$/hBN/tBLG are contacted by a shadow mask evaporation of Au at 300\,K. Afterwards the sample is transferred into a home-built ultrahigh vacuum scanning tunneling microscope (STM) operating at 300\,K \cite{Geringer2009} without further treatment. The instrument is equipped with a long-distance microscope to find flakes within the STM. W-tips, that are prepared on Au(111) priorily, are employed for STM images of the tBLG at sample voltage $V=300$\,mV and current $I=0.9$\,nA. The twist angle has been determined from the moiré periodicity using the simultaneously measured atomic periodicity as calibration.

\begin{figure}[h!]
    \centering
    \includegraphics{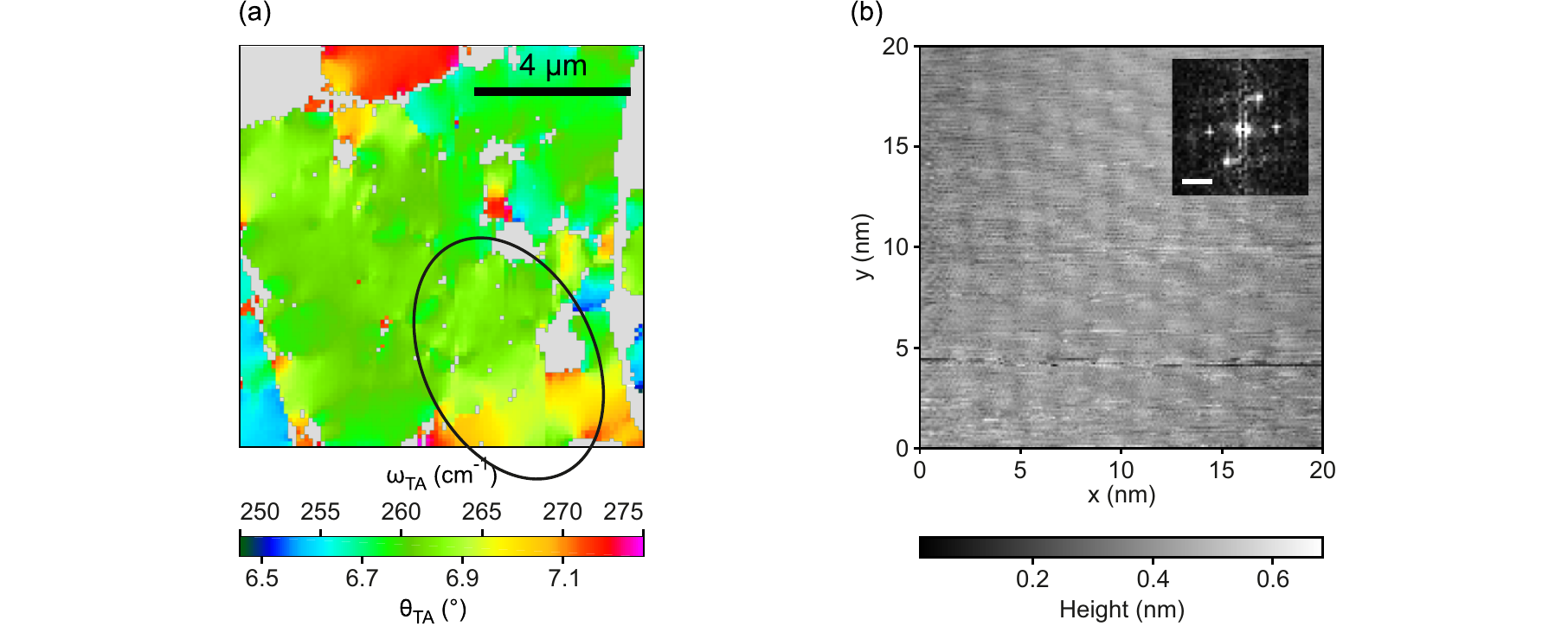}
    \caption{Comparison between Raman and STM measurements. (a) Zoom-in of the Raman map in Fig.~2e. The black ellipse indicates the approximate area where the STM measurement is taken. (b) STM map taken within the indicated region. The moiré wavelength is \SI{1.91 \pm 0.10}{\nano\meter}, corresponding to a twist angle of \SI{7.39 \pm 0.40}{\degree}.  The inset shows a Fourier transformation of the STM map with a scale bar of \SI{0.5}{\per\nano\meter}.}
    \label{fig:figS2}
\end{figure}
\subsection{Comparison to Raman experiment}
Figure \ref{fig:figS2}a shows a zoom-in of the Raman map shown in Figure~2e of the main text. It is difficult to accurately know the position where the STM map is recorded. Therefore, we marked the area after the measurement by piercing it with the STM tip, resulting in optically visible damage to the sample. The position and shape of the damaged area is marked by the ellipse in Figure~\ref{fig:figS2}a. Within this region, $\wTA$ varies between 253 cm$^{-1}$ and 271 cm$^{-1}$. Using our model as described in the main text, this results in twist angles between 6.76$^{\circ}$ and 7.24$^{\circ}$.

The STM map is shown in Figure~\ref{fig:figS2}b. From the Fourier transformation (inset), we estimate the moiré wavelength to be \SI{1.91 \pm 0.1}{\nano\meter}, resulting in a twist angle of \SI{7.39 \pm 0.4}{\degree}~\cite{Carozo2011Nov}. This confirms that the twist angle $\theta_{\mathrm{TA}}$ determined by Raman spectroscopy is close to the actual twist angle $\theta$ present in the sample.

\section{4. Discussion on inaccuracy in the twist angle estimation}
Here, we discuss the accuracy of our method to determine the absolute twist angle. The association of the frequency of the activated TA mode with a particular twist angle $\theta$ hinges on several assumptions or approximations of the phonon dispersion. We identified several effects that may lead to an inaccuracy of the twist angle estimation that we address below.

\begin{figure}[h!]
    \centering
    \includegraphics{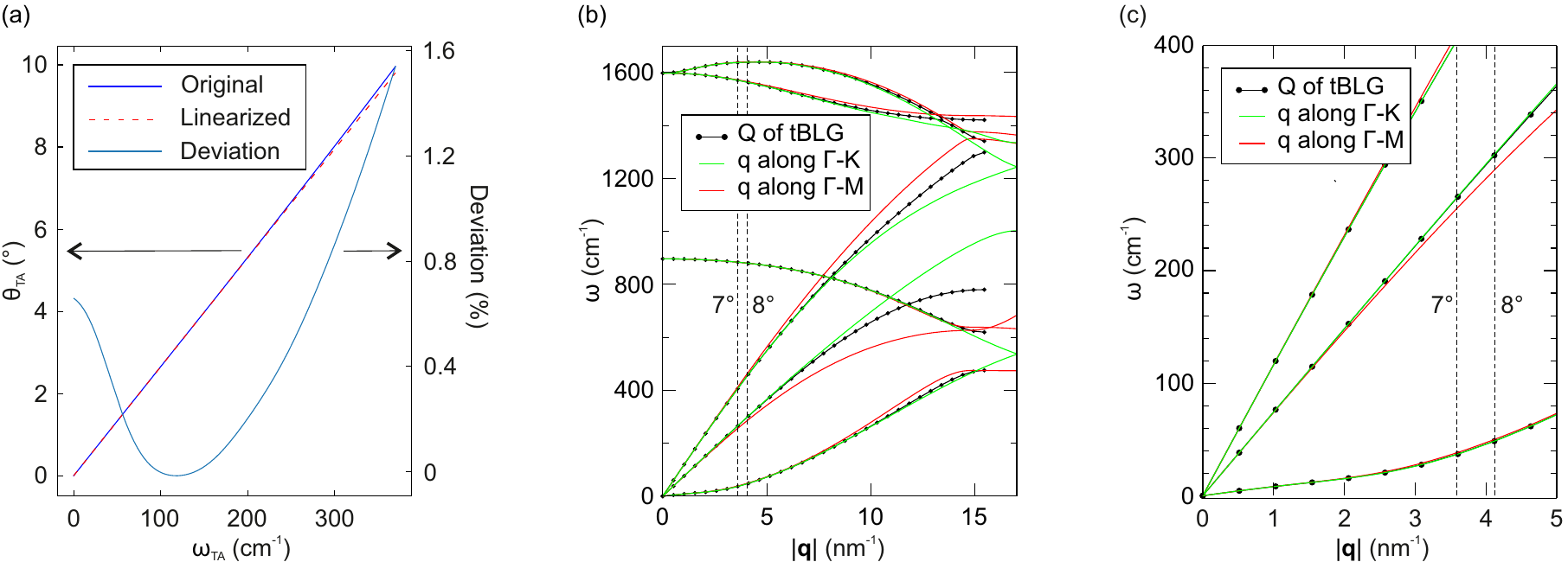}
    \caption{Details on the sources of systematic inaccuracy in twist angle calculation. (a) The effect of linearizing the TA phonon branch is illustrated by comparing the transformation of TA peak position $\omega_\mathrm{TA}$ into twist angle $\theta_\mathrm{TA}$ using the original dispersion (dark blue line) and the linearized dispersion (dashed red line). The light blue line shows the deviation between the two methods. (b) SLG band structure calculation along high-symmetry axes $\Gamma$-M, $\Gamma$-K and along the lattice vector of the mini-Brillouin zone. (c) Zoom-in into the calculation shown in panel (b) showing the range of crystal momentum $|\mathbf{q}|$ corresponding to twist angles from \SI{0}{\degree} to $\sim \SI{10}{\degree}$. }
    \label{fig:figS3}
\end{figure}

\subsection{Effect of the band dispersion linearization}
In the frequency range relevant for the analysis of low and intermediate twist angles, the TA phonon branch of graphene is almost linear.
As described in the main text, the phonon branch is linearized for the sake of simplicity in calculation.
In Figure~\ref{fig:figS3}a, we show how the linearization of the phonon dispersion affects the transformation of TA peak position $\omega_\mathrm{TA}$ into twist angle $\theta_\mathrm{TA}$ by comparing it to a transformation that was done using the original phonon dispersion.
The maximal deviation is $\sim \SI{1.5}{\percent}$ for twist angles $\theta < \SI{10}{\degree}$, as stated in the main text.

\subsection{Effect of the phonon dispersion calculation}
The translation of the TA peak position into the twist angle depends sensitively on the TA phonon dispersion.
The calculation of the phonon dispersion was performed as in Ref.~\cite{WirtzRubio04}, using density functional theory in the local-density approximation (LDA) for the exchange-correlation functional. Using the generalized-gradient approximation (GGA) yields almost exactly the same TA phonon dispersion (albeit the GGA lattice constant is 0.3\% larger than the LDA one). The TA mode at K is at 1001.3 cm$^{-1}$ in LDA and at 1001.8 cm$^{-1}$ in GGA\cite{WirtzRubio04} which means that the slope of the TA mode differs by less than 0.1\% between the two approximations.

One may wonder about the effect of layer-layer interaction on the slope of the TA mode. Here, we use the difference in the TA mode frequency at K between graphene (1001.3 cm$^{-1}$) and graphite (1001.9 cm$^{-1}$) which again gives a negligible impact on the slope of the TA branch (compared to the other sources of uncertainty discussed below).

\subsection{Effect of substrate and strain}
We expect a larger influence to come from the graphene substrate interaction via doping and/or strain effects. To estimate the impact of this effect, we consider the sound velocity of the TA phonon branch, $v_{\mathrm{TA}}$,
in the $\Gamma$-M direction.
From our phonon dispersion, we find $v_{\mathrm{TA}}^{\mathrm{theo}} = \SI{13.8}{\kilo\meter/\second}$.
while a recent experimental study of graphene on SiO$_2$, using resonant Raman spectroscopy, determined a value of $v_{\mathrm{TA}}^{\mathrm{exp}} = \SI{12.9}{\kilo\meter/\second}$ \cite{cong2019probing}.
The difference in slope by about 9\% may be taken as a measure for the error bar that gives an indication by about how much the real TA mode slope deviates from the ab-initio value.

A similar error bar for experimental TA slopes can be obtained by looking at the comparison of electron energy loss spectroscopy (EELS) dispersions \cite{Oshima81,Siebentritt97,Yanagisawa05} of graphene on different substrates with ab-initio calculations~\cite{WirtzRubio04}.

\subsection{Effect of the moiré lattice vector direction}
Another simplification consists in using the phonon dispersion along the high-symmetry axis $\Gamma$-K to transform the position of the TA peak into the length
\begin{equation}
Q(\theta) = |\mathbf{b}_{1,2}^\mathrm{s}(\theta)|=\frac{8\pi}{\sqrt{3}a}\sin\left(\frac{\theta}{2}\right)
\end{equation}
of the lattice vector of the mini-Brillouin zone (eq. 1 in the main text).
This would be true only if $\mathbf{b}_{1,2}^{\mathrm{s}}(\theta)$ always pointed along the $\Gamma$-K axis. This is approximately true for very small twist angles, but, in principle, both the length and the direction of these vectors depend on the twist angle.
In terms of reciprocal unit vectors $\hat{\mathbf{k}}_x$ and $\hat{\mathbf{k}}_y$, they can be expressed as~\cite{Carozo2011Nov}
\begin{equation}
    \mathbf{b}_{1,2}^{\mathrm{s}}(\theta) = \frac{2\pi}{\sqrt{3}a} \left[ [\mp(1-\mathrm{cos}(\theta)) - \sqrt{3}\:\mathrm{sin}(\theta)] \hat{\mathbf{k}}_x +
    [-\sqrt{3} (1-\mathrm{cos}(\theta)) \pm \mathrm{sin}(\theta)] \hat{\mathbf{k}}_y
    \right].
    \label{eq:eqs1}
\end{equation}
To confirm that the simplification to follow the $\Gamma$-K direction is valid nonetheless, we present in Figs.~\ref{fig:figS3}b-c a calculation of the phonon dispersion along the path defined in reciprocal space by eq.~\ref{eq:eqs1} for twist angles from \SI{0}{\degree} to \SI{30}{\degree}. For small twist angles, this path starts along the $\Gamma$-K direction, but then bends away from it and ends, for $\theta = $\SI{30}{\degree}, half way between K and M on the edge of the first Brillouin zone. We compare the phonon dispersion along this line to the phonon dispersion calculated along the high-symmetry axes $\Gamma$-K and $\Gamma$-M.
The dashed vertical lines denote twist angles of \SI{7}{\degree} and \SI{8}{\degree}.
In the regime of the twist angles that are relevant for this work, the dispersion along the path defined by eq.~\ref{eq:eqs1} lies very close to the dispersion in $\Gamma$-K direction.
Namely, for the TA mode at the wave vector corresponding to \SI{8}{\degree}, the deviation in frequency is \SI{0.85}{\per\centi\meter}, which corresponds to a deviation in twist angle of \SI{0.02}{\degree}.

\subsection{Inaccuracy in twist angle determination}
Based on the analysis of the different sources of uncertainty above, we find that in our twist angle range the uncertainty from the effect of substrate and strain significantly larger than the others. We thus take the 9\% uncertainty found here as the uncertainty in the dispersion that we use to determine our inaccuracy and precision in the main text. As stated in the main text, this uncertainty in the slope tranlates into an uncertainty of the \emph{absolute} value of the twist angle, but only marginally affects the experimental determination of twist angle \emph{variations}.

\end{document}